\documentclass[usenatbib]{mn2e}
\usepackage[utf8]{inputenc}
\usepackage{natbib,epsfig,cite,mystyle,amsmath,amssymb,graphicx}
\usepackage{array, xcolor, caption}
\usepackage[colorlinks=true,linkcolor=blue,citecolor=blue]{hyperref}

\title{Evolution of dispersion in the cosmic deuterium abundance}
\author[I. Dvorkin, E. Vangioni, J. Silk, P. Petitjean, K. Olive]{Irina
Dvorkin$^{1}$\thanks{E-mail: dvorkin@iap.fr}, Elisabeth Vangioni$^{1}$, Joseph Silk$^{1,2}$, Patrick
Petitjean$^{1}$, Keith A. Olive$^{3}$\\
$^{1}$Sorbonne Universités, UPMC Univ Paris 6 et CNRS, UMR 7095, Institut
d’Astrophysique de Paris, 98 bis bd Arago, 75014 Paris, France \\
$^{2}$Department of Physics and Astronomy, The Johns Hopkins University,
Baltimore, MD 21218, USA \\
$^{3}$William I. Fine Theoretical Physics Institute, School of Physics and
Astronomy, University of Minnesota, Minneapolis, MN 55455, USA
}
\begin{document}

\pagerange{\pageref{firstpage}--\pageref{lastpage}} \pubyear{2016}
\maketitle
\label{firstpage}

\begin{abstract}

Deuterium is created during Big Bang Nucleosynthesis, and, in contrast to the
other light stable nuclei, can only be destroyed thereafter by fusion in stellar
interiors. In this paper we study the
cosmic evolution of the deuterium abundance in the interstellar medium and its dispersion using realistic galaxy evolution
models. We find that models that reproduce the observed metal abundance are
compatible with observations of the deuterium abundance in the local ISM and
$z\sim 3$ absorption line systems. In particular, we reproduce the low astration
factor which we attribute to a low global star formation efficiency. We
calculate the dispersion in deuterium abundance arising from different structure
formation histories in different parts of the Universe. Our model also predicts
a tight correlation between deuterium and
metal abundances which could be used to measure the primordial deuterium abundance. 
\end{abstract}

\section{Introduction}

Deuterium is created during Big Bang Nucleosynthesis (BBN) and can only be destroyed thereafter as its fusion temperature is of the
order of $10^5$ K \citep{1976Natur.263..198E}. Therefore, any detection of
deuterium sets a lower limit on its primordial value $(D/H)_p$ and can be used
to constrain the baryon-to-photon ratio $\eta$. 

The conversion from $\eta$, provided by CMB
measurements, to $(D/H)_p$ is not straightforward in view of the uncertainties
in the nuclear cross sections. Although minute, these errors are becoming
increasingly important as the quality of data increases. In this paper we
consider two recent calculations: \citet{2015arXiv151103843C}, which results in
$10^5(D/H)_p=2.45\pm 0.05$ for $10^{10}\eta=6.09$ and \citet{2015arXiv150501076C}
with $10^5(D/H)_p=2.58\pm 0.13$ for $10^{10}\eta=6.10$. We stress that both
results are compatible, within the quoted errors, with each other and with the
abundance measured by \citet{2014ApJ...781...31C}, as well as with the value of $10^5(D/H)_p=2.49\pm 0.03\pm 0.03$ deduced by \citet{2015arXiv151007877M}.

Deuterium is detected in several metal-poor damped Ly$\alpha$ absorbers (DLAs)
where $(D/H)$ is expected to be close to its primordial value.
Early reported values of $D/H$ 
\citep{bt98a,bt98b,omeara,pettini,lev,kirkman,omeara2,pettini2,sr10,2011Sci...334.1245F} 
showed a high degree of dispersion
which was significantly larger than the
quoted errors. This dispersion may have been intrinsic \citep{2001ApJ...563..653F} or
most probably systematic.
Following the observation of \citet{2012MNRAS.425.2477P},
\citet{2014ApJ...781...31C} proposed to circumvent this difficulty by defining a
\emph{Precision Sample} of deuterium abundance measurements, which adhere to a
strict set of selection criteria, thus ensuring its robustness. This sample,
which consists of $5$ points at $z\sim 3$, provides a mean weighted value of the
primordial abundance of $10^5(D/H)_p=2.53\pm 0.04$. Subsequently two new obervations
\citep{2015MNRAS.447.2925R,2015arXiv151101797B} were added to the literature
albeit with significantly larger uncertainties.

Another set of observational constraints comes from local measurements,
including Jupiter, interstellar matter (ISM) in the solar neighbourhood, the galactic disk and
the warm
neutral medium in the galactic halo \citep[see][for a
compilation of results]{2007ApJ...659.1222S}. Since deuterium is destroyed as gas is cycled through stars, the astration factor $f_a=(D/H)_p/(D/H)\geq 1$ is a measure of the star formation and inflow history in the Galaxy. Unfortunately, local observations also suffer from large dispersion, partly due to the variance in dust depletion along different lines of sight \citep{2006ApJ...647.1106L}. Overall the results imply an astration factor in the range $f_a\simeq
1.1-1.4$, reflecting relatively low star formation efficiency.

The chemical evolution of D/H has been extensively studied \citep[see e.g.][]{1974ApJ...192..487A,
vangioni88,1992ApJ...401..150S,1994ApJ...427..618V,1997ApJ...476..521S,2006MNRAS.369..295R,2007MNRAS.378..576S,
2012MNRAS.426.1427O}.
In this paper we study the evolution of deuterium abundance in a full cosmological context
and test to what extent the abundance observed at different redshifts can
be thought of representing the primordial value.
Using realistic galaxy evolution models we study the connection between
deuterium and metal abundances, which provide complementary views on the star
formation process. We
show that models that succesfully explain the observed total mean metallicity
abundance at high redshift are consistent with $(D/H)$ measurements in $z\sim
3$ DLAs and the local ISM and predict a significant dispersion in $(D/H)$ which
evolves with redshift. Finally, we predict a tight correlation
between the deuterium and metal content in the ISM, which can be exploited for
extending the range of systems used to
measure $(D/H)_p$.

 In section \ref{sec:model} we review
our model of galaxy evolution. In section \ref{sec:deu} we show the resulting
deuterium abundance in the ISM
and discuss the evolution of its dispersion. Conclusions are presented in
section \ref{sec:disc}.

\section{Galaxy evolution model}
\label{sec:model}

We use an improved version of the galaxy evolution model developed
in \citet{2015MNRAS.452L..36D}. We
take $V_{tot}=10^6$ (Mpc/h)$^3$ as the total comoving volume in our
calculation and divide it into $1000$ smaller \emph{regions} $\Delta V_i=10^3$
(Mpc/h)$^3$. The volume $V_{tot}$ is populated with dark matter
(DM) halos according to the
Sheth-Tormen mass function (MF)
at $z=0$ accounting for large-scale clustering effects: regions with large-scale overdensities form halos more easily. We use the results of \citet{2004ApJ...609..474B}, for 
the MF bias caused by clustering (see their eqs. (6,7)). This
correction provides a good fit to numerical simulations at least up to $z\simeq
30$. For each region, we draw an overdensity $\delta_i$ from a Gaussian distribution with
zero mean and variance $\sigma(R=10$ Mpc$)$, where $\sigma(R)$ is the variance of the overdensity field smoothed on a scale $R$ and employ
the formulas from \citet{2004ApJ...609..474B} to obtain the halo MF in the
region. The actual mass distribution in each region is obtained by drawing
halos from this biased MF. This procedure is equivalent to smoothing the real halo distribution on a scale of $R=10$ Mpc, roughly corresponding to the galaxy correlation length. 

We build a merger tree for each halo using the algorithm in the
GALFORM model \citep{2008MNRAS.383..557P} and follow its evolution
backwards in time up to $z_{f}=15$, saving the output in $50$ equally spaced redshift bins. 
We use $M_{min}=10^8M_{\odot}$ as the minimal halo
mass able to form stars. In this manner, we obtain the distribution of DM halos
in the comoving volume $V_{tot}$ as a function of redshift.

We then calculate the mean mass fraction in collapsed structures $f_{coll,i}$
and
the mean
escape velocity, $v_{esc}$, in each region $i$. These quantities serve as inputs for the
chemical evolution code developed in
\citet{2004ApJ...617..693D,2006ApJ...647..773D} and \citet{2009MNRAS.398.1782R},
which follows the exchange of
mass between the gas within and outside of collapsed structures, the star formation rate (SFR) at each
redshift and the rate of metal production in stars. The evolution of gas mass in
collapsed structures is given by:
\begin{equation}
 \dot{M}_{ISM}=a_b(t)-o(t)-\psi(t)+e(t)
\label{eq:mdot}
\end{equation}
where $\psi(t)$ is the SFR, $a_b(t)$ is the rate of gas accretion, $e(t)$ is
the rate at which processed gas is returned to the ISM following stellar deaths
and $o(t)$ is the outflow rate. In our model $a_b(t)$ is calculated from
$f_{coll}(t)$ for every region, and the other terms in eq. (\ref{eq:mdot})
depend explicitly on the SFR. 

\begin{figure}
\centering
\epsfig{file=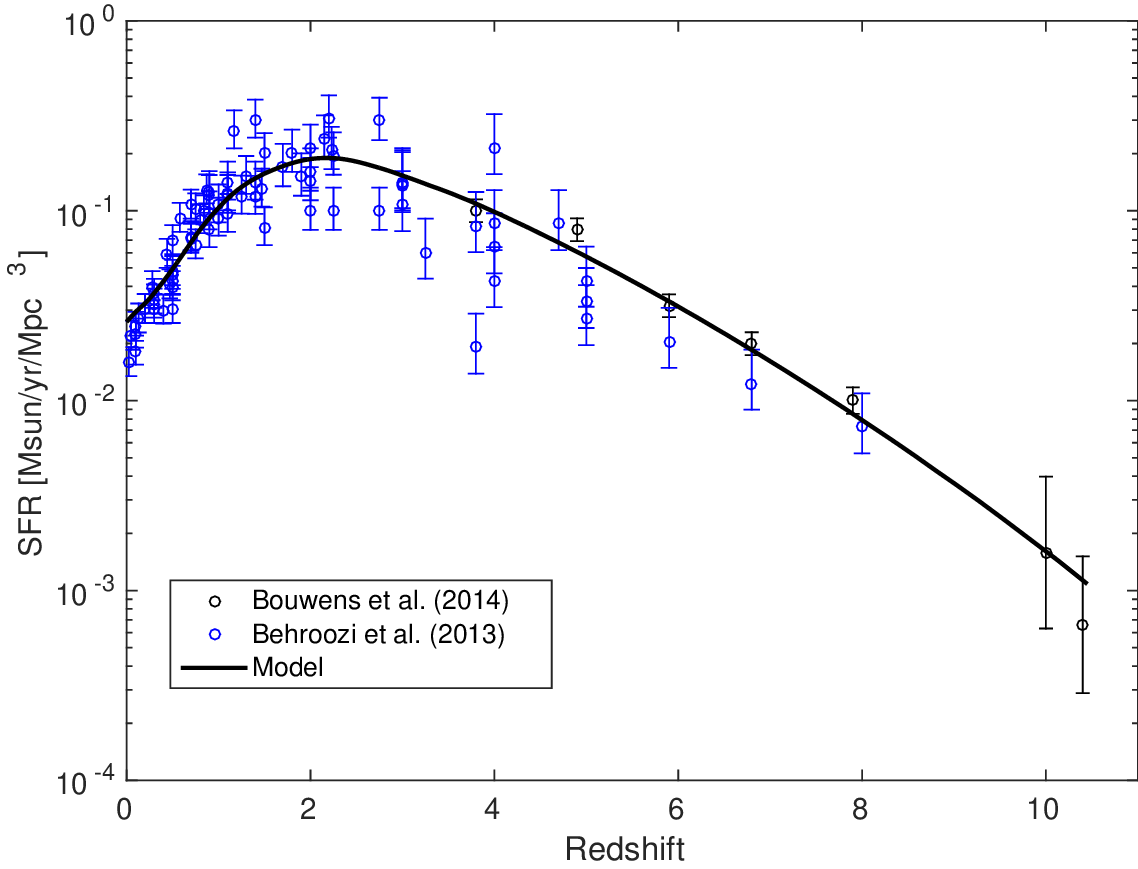, height=5.2cm}
\caption{The mean model SFR (black line) compared with observations
taken from \citet[][]{2013ApJ...770...57B} (blue points) and \citet{2014ApJ...795..126B} black points.}
\label{fig:sfr10}
\end{figure}

The mass of stars formed in a given galaxy clearly depends on the amount of
available baryonic fuel, which is determined by inflow and outflow processes.
Therefore we use a phenomenological model where the SFR depends on the mass of
the ISM at each time-step of the calculation: 

\begin{equation}
\psi= \begin{cases}
\nu\exp\left[-(\log M_{ISM}-\log\bar{M}_{p})^2/\sigma_1^2\right] &\mbox{if }
M_{ISM} \geq  \bar{M}_{p} \\
\nu\exp\left[(\log M_{ISM}-\log\bar{M}_{p})/\sigma_2\right]&
\mbox{if } M_{ISM} < \bar{M}_{p}.
  \end{cases}
\end{equation}
where $\nu=0.23$ M$_{\odot}$ yr$^{-1}$ Mpc$^{-3}$, $\bar{M}_p=4.3\times 10^{10}
M_{\odot}$, $\sigma_1=0.045$ and $\sigma_2=0.24$. In this model star formation
is quenched as soon as the galaxy acquires a critical mass of $\bar{M}_{p}$. 
The mean cosmic SFR is shown in
Figure \ref{fig:sfr10} (black solid line) and compared with observations
compiled
by \citet{2013ApJ...770...57B} and the data from \citet{2014ApJ...795..126B}. 

The calculation of outflow rate is performed as in
\citet{2006ApJ...647..773D} and is given by: 
\begin{equation}
 o(t)=\frac{2\epsilon}{v_{esc}^2(t)}\int dm \Phi(m) \psi(t-\tau(m)) E_{kin}(m)\:,
\end{equation}
where $\tau(m)$ is the lifetime of a star with mass $m$, $\Phi(m)$ is the
stellar initial mass function (IMF), $E_{kin}$ is the
kinetic energy released when this star dies, and $\epsilon=0.001$ is the fraction of
kinetic energy that powers the outflow. This value of $\epsilon$ was chosen by comparing the IGM metallicities predicted by the model and the observed oxygen and carbon abundances in the Ly-alpha forest \citep[see discussion in][]{2006ApJ...647..773D}. We assume a Salpeter IMF with slope $-2.35$ for $0.1 \leq m/M_{\odot} \leq 100$. Even though this choice differs from more recent IMFs \citep[e.g][]{2001MNRAS.322..231K}, notably in the low-mass end, we do not expect this to be important for deuterium abundance evolution. The effect of different IMFs on the mean deuterium abundance was discussed in \citet{2015arXiv151103843C} and it was found that taking $0.5M_{\odot}$ as the lower mass limit reduces the deuterium abundance at $z=0$ by less than $10\%$, and the effect is even smaller at higher redshifts. Similarly, modifying the IMF slope to $-2.7$ \citep{2014ApJ...796...75C} amounts to $\lesssim 5\%$ change in deuterium abundance.

The rate at which processed gas is returned to the ISM is given by:
\begin{equation}
 e(t)=\int \Phi(m)\psi(t-\tau(m))(m-m_r)dm\:,
\label{eq:et}
\end{equation}
where $m_r$ is the remnant mass of a star of mass $m$.

The evolution of the abundance of a specific element $i$ in the ISM and the
intergalactic medium (IGM) is
given by eqs. (6) and (7) in \citet{2004ApJ...617..693D}. For the sake of
completeness we reproduce the equation for the ISM:
\begin{equation}
 \dot{X}_{i}^{ISM}=\frac{1}{M_{ISM}(t)}\left\lbrace e_i(t)-e(t)X_i^{ISM}
+a_b(t)\left[X_i^{IGM}-X_i^{ISM} \right]\right\rbrace
\label{eq:dxdt}
\end{equation}
where $e_i(t)$ is the rate at which element $i$ is ejected into the ISM and is
given by:
\begin{equation}
 e_i(t)=\int \phi(m)\psi(t-\tau(m))m_{ej}^i(m)dm
\label{eq:eit}
\end{equation}
and $m_{ej}^i(m)$ is computed from stellar yields.

\section{Deuterium abundance}
\label{sec:deu}

\begin{figure}
\centering
\epsfig{file=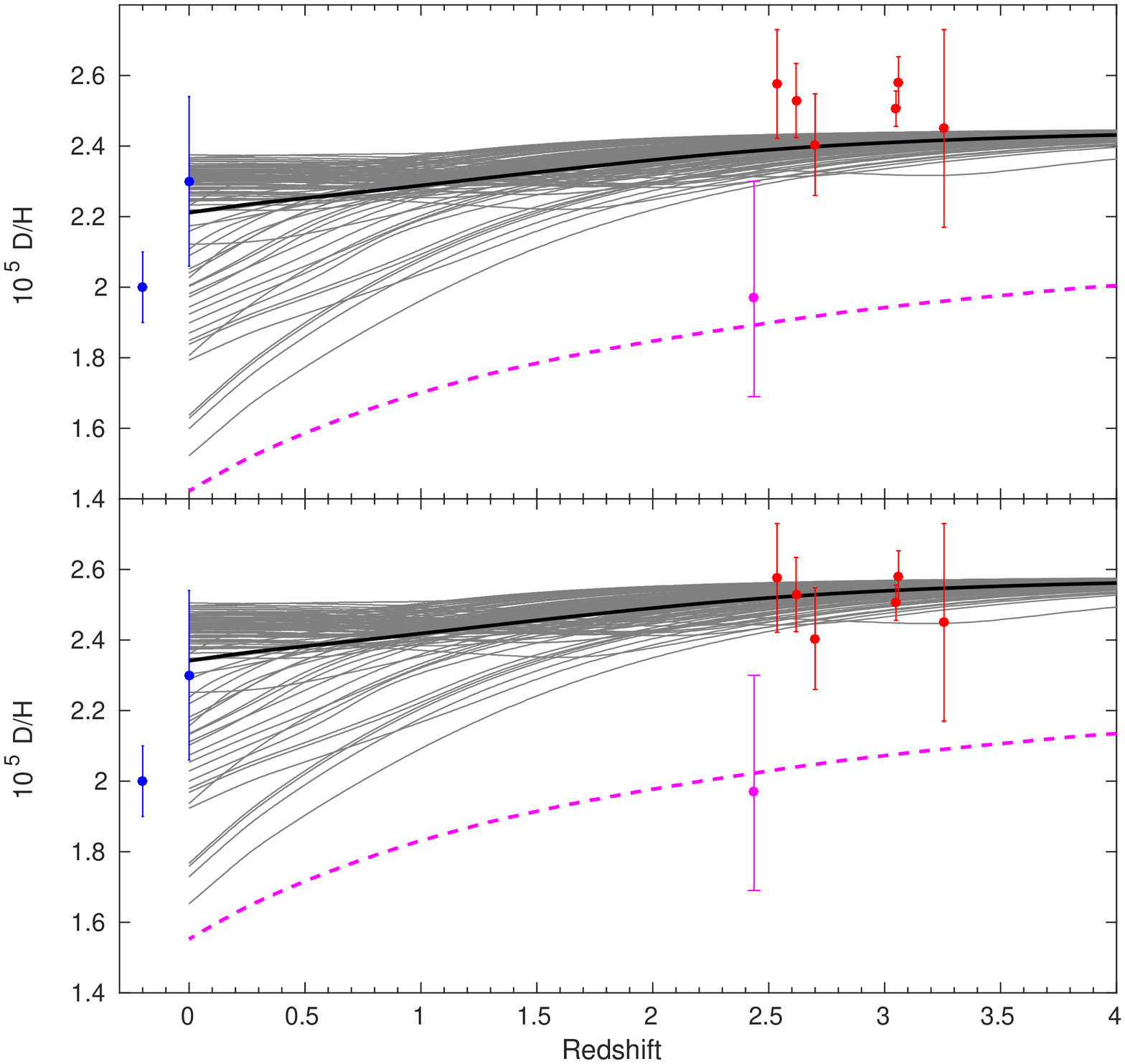, height=8cm}
\caption{The evolution of deuterium abundance in each region (thin grey lines), the mean (black line) and the case with PopIII stars (dashed magenta). 
The sample  of \citet{2014ApJ...781...31C} and \citet{2015MNRAS.447.2925R} 
is shown in red, that of \citet{2015arXiv151101797B} in magenta, and local ISM measurements \citep{2006ApJ...647.1106L,2010MNRAS.406.1108P} in blue (one of the points at $z=0$ is shifted for clarity). \emph{Upper panel:} primordial value of
$10^5(D/H)_p=2.45$. \emph{Lower panel:} primordial value of
$10^5(D/H)_p=2.58$.}
\label{fig:meand11_sel}
\end{figure}

The evolution of deuterium abundance in the ISM is determined by the star
formation history, the rate of stellar deaths, which return processed gas into
the ISM and the inflow of primordial gas into the galaxy. We therefore expect
different regions in the Universe to have different mean levels of deuterium
abundance, according to their structure formation history.
The mean evolution of deuterium abundance in our model assuming the primordial values of $10^5(D/H)_p=2.45$ \citep{2015arXiv151103843C} and $10^5(D/H)_p=2.58$ \citep{2015arXiv150501076C} is shown by the black lines on Figure
\ref{fig:meand11_sel} (upper and lower panels, respectively). The grey lines show the evolution in $100$ individual regions in our model,
as discussed below. The abundances measured in DLAs at $z\sim 3$ from
 \citet{2014ApJ...781...31C} and \citet{2015MNRAS.447.2925R} are shown in red and the measurement by \citet{2015arXiv151101797B} is shown as magenta. In addition, we show the
deuterium abundance measured in the local ISM
\citep{2006ApJ...647.1106L,2010MNRAS.406.1108P} (blue) (note that local measurements might have additional uncertainties due to dust depletion effects, see e.g. \citet{2003ApJ...599..297H}. 

Comparing the two theoretically deduced values for $(D/H)_p$ taken from the recent studies of \citep{2015arXiv151103843C} and \citep{2015arXiv150501076C}, we see that, as expected, there is no qualitative change in the
results, except for a shift of the whole distribution.

We now turn to a more detailed analysis of the dispersion in the deuterium abundance by inspecting the individual regions shown by thin grey lines on Figure \ref{fig:meand11_sel}. Note that there are regions where the deuterium abundance reaches a minimum, then
begins to rise. This effect occurs in regions where star formation was quenched
early on so that deuterium is no longer destroyed. However if subsequent
accretion of primordial material is significant, the deuterium abundance in the depleted
ISM will increase. 

Comparing with the available measurements in $z\sim 3$ DLAs and in the local ISM, we see that our model is consistent with the data of \citet{2014ApJ...781...31C} and \citet{2015MNRAS.447.2925R}, as well as with the local measurements. A few regions in our model also agree, within the quoted uncertainty, with the low deuterium abundance found by \citet{2015arXiv151101797B} for a $z=2.4$ absorption system.  However, it is also possible that the true abundance of J$1444+2919$ is much lower than any of our models predict. In the most extreme star-forming region in our model, deuterium is depleted to $\sim 1/3$ of its initial value at $z=0$, but at $z=2.4$ its abundance is still slightly above $2.2\times 10^{-5}$. This
result is related to the relative inefficiency of star formation, as most
material never went through a star formation phase. Standard theories of galaxy
formation will therefore have difficulties explaining the extremely low
deuterium abundance of J$1444+2919$, in particular in view of its low metal
abundance of $[O/H]=-2.04$. An interesting scenario is an early star formation
episode that could have led to local depletion of deuterium. The dashed magenta
curve on Figure \ref{fig:meand11_sel} shows one such model with an addition of
PopIII stars for which the SFR is $\psi_p=\nu_p
M_{ISM}\exp(-Z_{IGM}/Z_{crit})$, where $Z_{IGM}$ is the metallicity of the IGM,
$\nu_p=2$ Gyr$^{-1}$ and we assumed a rather high critical metallicity of
$Z_{crit}=0.01$. This model results in early depletion of deuterium in the
concerned region and is able to explain the low deuterium abundance reported by
\citet{2015arXiv151101797B}. However this model also predicts a relatively high
metal abundance of $[O/H]=-0.5$. A possible solution would be to include strong
outflows by supernovae ejecta that preferentially remove metals (through the chimney effect \citep{snow})
while preserving
the low abundance of deuterium. Note, however, that $HI$ column density in the
analysis of \citet{2015arXiv151101797B} could be overestimated, yielding higher
values of both $(D/H)$ and metallicity and bringing this point into the range
predicted by our models.

\begin{figure}
\centering
\epsfig{file=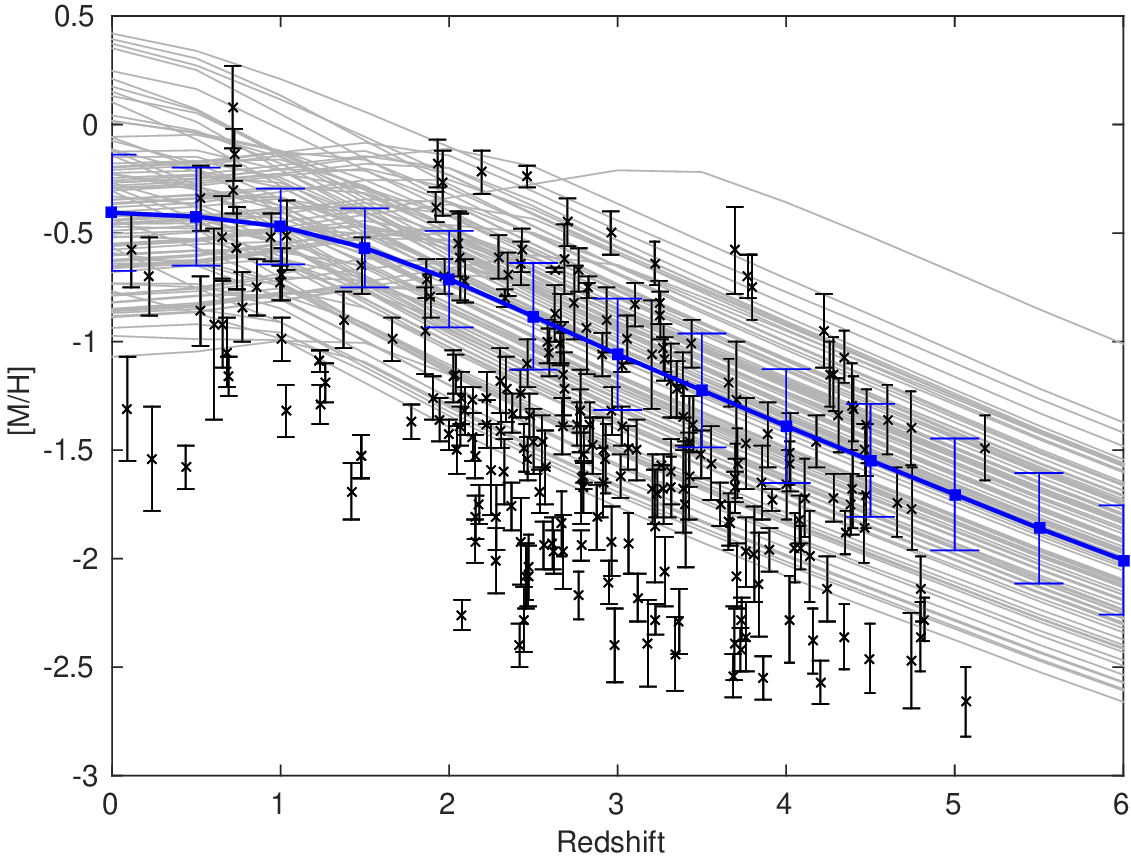, height=5.2cm}
\caption{The dispersion in metallicity abundance in our model (thin grey lines)
and the mean (blue solid line) compared with DLA measurements from \citet{2012ApJ...755...89R}. 
The blue points show the variance about the mean metallicity. Our models explain the bulk of the measurements, with the exception of extremely metal-poor systems (see text).}
\label{fig:metals11}
\end{figure}

We also note that the
dispersion in deuterium abundance is significantly smaller than the expected
metallicity dispersion, shown in Figure \ref{fig:metals11}, which spans several
orders of magnitude. This is expected since both effects are the result of
differences in the amount of matter that was cycled through stars, however in the
case of metals the dispersion was effectively amplified by the stellar
yields. We note that our model fails to account for
the most metal-poor systems and thus underestimates the total dispersion. A
complete treatment of this problem, including the effects of the structure of
the galactic halo is beyond the scope of the present paper and we leave it to
future work.

\begin{figure}
\centering
\epsfig{file=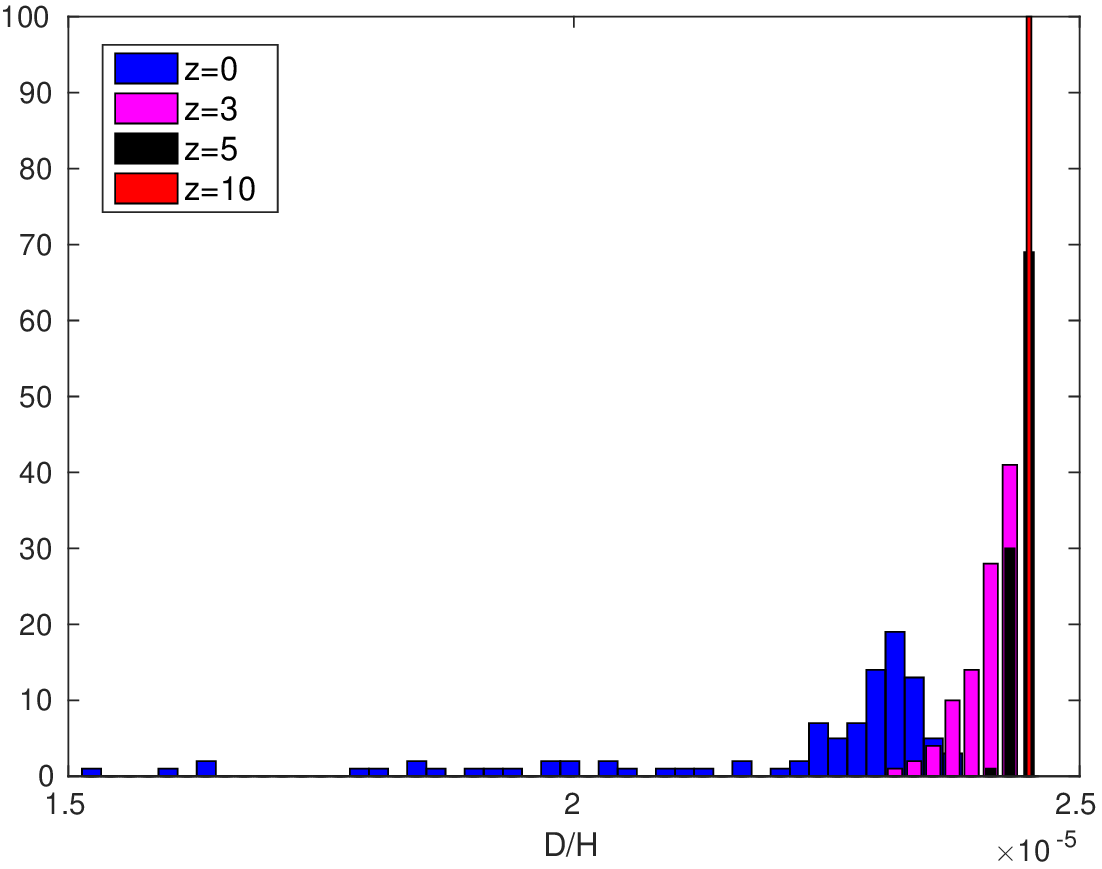, height=5.2cm}
\caption{The histogram of deuterium abundance at $4$ different redshifts for the
primordial value of $10^5(D/H)_p=2.45$.}
\label{fig:histd10}
\end{figure}

To better understand the dispersion of deuterium abundance, 
in Figure \ref{fig:histd10}, we construct
histograms of $(D/H)$ at different redshifts, assuming a primordial abundance of $10^5(D/H)_p=2.45$. At $z=10$ the deuterium abundance is at the primordial level
everywhere, but it begins to slightly decline by $z=5$. At $z=3$ the
distribution is quite wide, so that more precise measurements of DLA absorption
might detect a significant deviation from the primordial value and reveal the
star formation history of these systems. Interestingly, at $z=0$ the
distribution consists of a Gaussian-like feature around $10^5(D/H)=2.3$ and a long low-abundance tail, which corresponds to regions where star
formation continues at low redshifts. The evolution of dispersion with redshift
reflects the star formation history in our model and thus in principle can
constrain galaxy evolution models. Although current observational data does not
provide sufficient statistics, we expect future surveys of absorption line
systems to provide interesting constraints on star formation histories of
high-redshift galaxies.

\begin{figure}
 \centering
 \epsfig{file=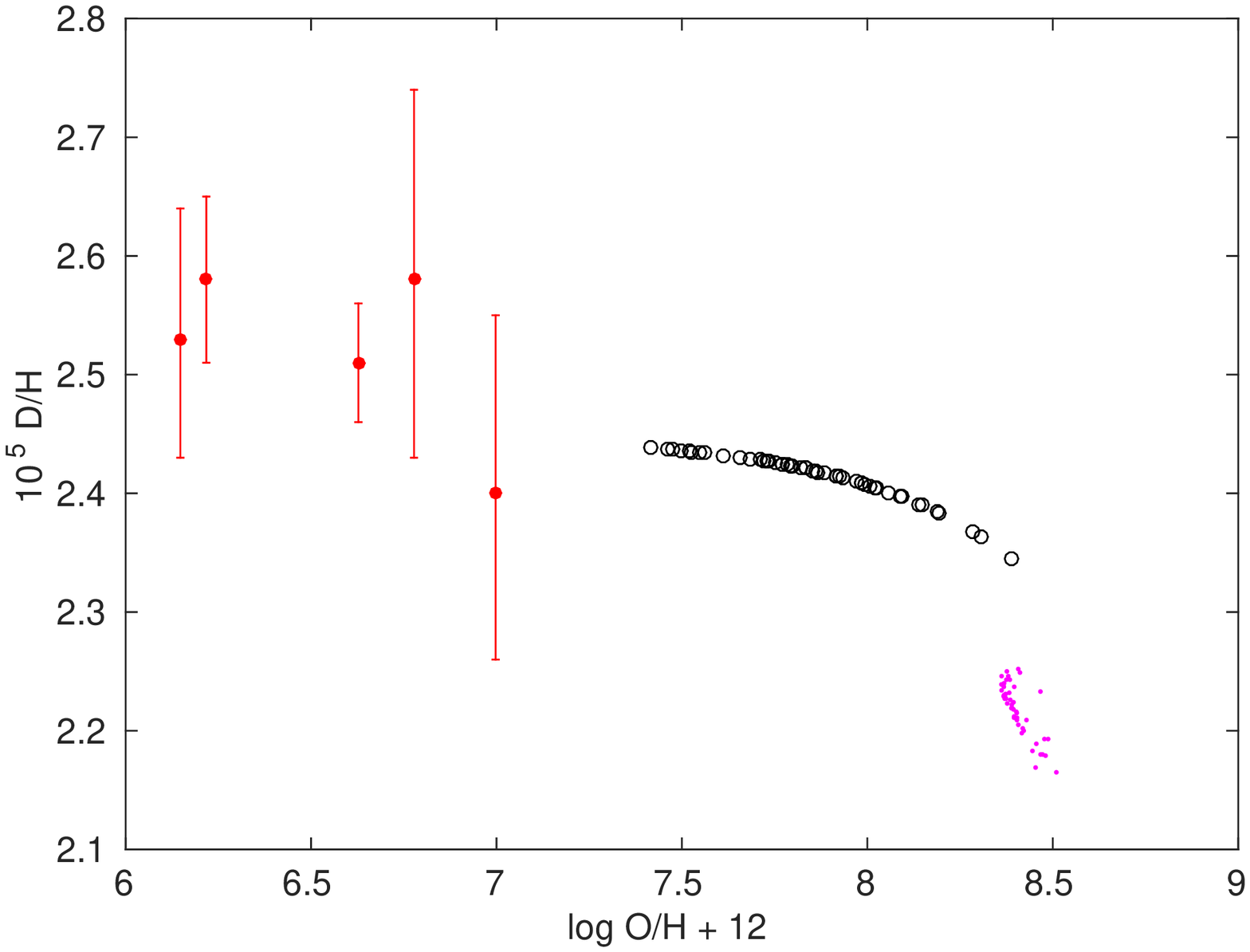, height=5cm}
 \caption{The correlation between $(D/H)$ and $(O/H)$ at $z=3$ for our fiducial model with $10^5(D/H)_p=2.45$ (black circles) and with an early episode of star formation (magenta dots) compared with observations from \citet{2014ApJ...781...31C} in the redshift range $2.4-3.1$ (red points). Our model does not describe the extremely metal-poor systems (compare also with Figure \ref{fig:metals11}), see text for discussion. The correlation between $(D/H)$ and metallicity can potentially be used to measure the primordial abundance.}
 \label{fig:DZ11}
 \end{figure}

Deuterium and metal abundances are complementary tracers
of the star formation history of a given region: star formation depletes
deuterium while increasing the fraction of metals. We can thus expect these two
quantities to be tightly correlated. This is in fact what happens in our model,
as can be seen in Figure \ref{fig:DZ11} (black circles). As noted above, our model
does not reproduce the most metal-poor systems, probably because it does not
resolve individual halos and their internal structure. However our
results clearly demonstrate the small amount of astration even in systems with
$\log (O/H)+12\simeq 7.5$ as well as the correlation between metallicity and
$(D/H)$. This suggests that DLAs with intermediate metallicity values might be
interesting in the context of measuring the primordial deuterium abundance.

The correlation between $(D/H)$ and metallicity can be
understood by considering our model of chemical evolution for deuterium and an
arbitrary element $i$. From eq.
(\ref{eq:dxdt}) and assuming negligible metal abundance in the IGM, negligible
astration: $X_{D}^{ISM}(t)=X_{D,p}$ and negligible enrichment of the
ISM: $X_i^{ISM}(t)(a_b(t)+e(t))\ll e_i(t)$ (which hold when star formation is
very inefficient) we obtain the ratio of the time derivative of deuterium
abundance in the ISM to the time derivative of the abundance of element $i$:
\begin{equation}
 \frac{\dot{X}_D^{ISM}}{\dot{X}_i^{ISM}}=-\frac{e(t)X_{D,p}}{e_i(t)}\:.
\end{equation}
As long as this ratio is roughly constant in time, $(D/H)$ and metallicity will
be related by the slope $-X_{D,p}/y_{i,eff}$ where the effective yield is
$y_{i,eff}=e_i(t)/e(t)$. However if one of the aforementioned assumptions is violated, the correlation is destroyed. For example, the magenta dots in Figure \ref{fig:DZ11} show the case with an early strong star formation episode with $\psi=50$ Gyr$^{-1}$ for $t< 0.5$ Gyr.

We stress that this theoretical relationship should be further explored, in
particular the extent to which our assumptions are valid in realistic systems
and the conclusions that can be drawn from a departure from this relation (i.e. a
non-standard star formation history). We plan to explore these issues in the
context of a more complete semi-analytic model that resolves individual halos,
to be presented in a forthcoming paper. 

\section{Discussion}
\label{sec:disc}

In this paper we developed a model that describes the evolution of the deuterium
abundance in the context of cosmological structure formation. Employing
realistic descriptions of the SFR and galactic outflows we were able to obtain a
good fit to the observed deuterium abundance in the \emph{Precision Sample} of \citet{2014ApJ...781...31C} at $z\sim 3$ as well as
in the local ISM. 

In addition, we calculated the disperison
expected from different structure formation histories. We find that even in the
most extreme cases deuterium abundance is never reduced below $\sim 1/3$ of its
primordial value. Moreover, we find that the evolution of
deuterium can be non-monotonic if matter accretion continues after the quenching
of star formation, however, as expected, its abundance in the ISM never exceeds
the primordial value. 

An interesting outcome of our model is the tight correlation between the
deuterium and metal abundances in any given region. 
This finding suggests that it
might be possible to compile
large samples from different environments in order to study the primordial
deuterium abundance, rather than targeting only metal-poor systems. We note,
however, that metal-rich systems are expected to contain large amounts of dust
which will introduce additional scatter.
According to our model, DLAs which are poor in deuterium are outstanding candidates to be further studied at high sensitivity for possible Population III abundance signatures, since in these systems deuterium was efficiently destroyed in parallel with metal enrichment.
More observational and theoretical work is
needed to improve our understanding 
of the relation between deuterium and metal
abundances, which can provide important constraints on galaxy formation models. 

\section*{Acknowledgements}                                                      
We thank the \textsc{galform} team for making the code
publicly available. The work of ID and JS was supported by the ERC Project No. 267117 (DARK) hosted by Universit\'{e} Pierre et Marie Curie (UPMC) - Paris 6, PI J. Silk. JS acknowledges the support of the JHU by NSF grant OIA-1124403. The work of KAO was supported in part by DOE grant DE-SC0011842 at the University of Minnesota. This work has been carried out at the ILP LABEX (under reference ANR-10-LABX-63) supported by French state funds managed by the ANR
within the Investissements d'Avenir programme under reference ANR-11-IDEX-0004-02.
\bibliographystyle{mn2e}

\end{document}